\documentclass[llncsdoc]{llncs}

\usepackage[T1]{fontenc}

\usepackage{llncsdoc}
\usepackage{rotating}
\usepackage{etoolbox}
\usepackage{array,url}
\usepackage{algpseudocode}
\usepackage{subfig}
\usepackage{algorithm}

\newcolumntype{R}[2]{%
	>{\adjustbox{angle=#1,lap=\width-(#2)}\bgroup}%
	l%
	<{\egroup}%
}
 \newsavebox\cellbox
 \newcolumntype{w}[2]{%
>{\begin{lrbox}\cellbox}%
	l%
<{\end{lrbox}%
\makebox[#2][#1]{\usebox\cellbox}}}

\newcolumntype{C}[1]{wc{#1}}
\newcolumntype{L}[1]{wl{#1}}
\newcolumntype{R}[1]{wr{#1}}
\newcolumntype{W}[2]
{>{\begin{lrbox}\cellbox}%
		l%
		<{\end{lrbox}%
		
		\makebox[#2][#1]{\unhbox\cellbox}}}

\newcolumntype{C}[1]{>{\raggedleft\arraybackslash}p{#1}}

\newtoggle{conf}
\toggletrue{conf}
\newtoggle{proof}
\togglefalse{proof}
\usepackage{color}
\usepackage{amsmath,amssymb}
\usepackage{graphicx}
\usepackage{subfig}
\usepackage{multirow}
\usepackage{eurosym}
\usepackage{wrapfig,booktabs}

\newcommand{\orig}[0]{\textnormal{\textsf{\bf {orig}}}}

\newcommand*{\LargerCdot}{\raisebox{-0.75ex}{\scalebox{2.5}{$\cdot$}}}

\newcommand{\THEN}[0]{{\textnormal{\bf then}}}

\newcommand{\help}[1]{{}}

\newcommand{\term}[1]{{\emph{#1}}}

\newcommand{\upto}[0]{{..}}

\def\caA{{\cal A}}

\def\caC{{\cal C}}
\def\caD{{\cal D}}
\def\caE{{\cal E}}

\def\caM{{\cal M}}
\def\caR{{\cal R}}
\def\caS{{\cal S}}
\def\caT{{\cal T}}
\def\caU{{\cal U}}

\def\Att{\caA}
\def\Data{\caD}
\def\Ev{\caE}
\def\EvC{\caE_{cmp}}
\def\EvE{\caE_{ext}}
\def\Task{\caT}
\def\Stg{\caS}
\def\Mst{\caM}
\def\Ud1{\caU_d}
\def\Ue1{\caU_E}
\def\US1{\caU_S}
\def\UM1{\caU_m}

\def\rules{\caR}
\def\rulesPlus{\rules_+}
\def\rulesMinus{\rules_-}

\newcommand{\GammaOne}[0]{\Gamma^{1}}
\newcommand{\GammaTwo}[0]{\Gamma^{2}}

\newcommand{\GammaN}[0]{\Gamma^{n}}
\newcommand{\GammaBase}[0]{\Gamma^{base}}

\def\after{\mathop{\LargerCdot}}

\newcommand{\cmpltn}[1]{\mbox{{\sf C:}}{#1}}  % as in ``completion''
\newcommand{\evt}[1]{\mbox{{\sf E:}}{#1}}  % as in ``event''

\newtheorem{deff}{Definition}
\begin{document}
\markboth{}{}
\title{Extracting Features From Process Variants in Case Management}

\author{Rik Eshuis}
\institute{Eindhoven University of Technology, P.O.\ Box 513, 5600 MB, The Netherlands
	\email{h.eshuis@tue.nl}}

\maketitle

\begin{abstract}
Case Management supports knowledge workers in performing knowledge-intensive processes in a flexible way.
An essential ingredient of Case Management are template processes that are modified for a specific case to suit the context of that case.
Modifying templates results in many different yet related process variants.
However, modifying a template is time consuming and may lead to errors.
This paper defines an approach to extract fragments, called features, from artifact-centric process variants in case management. By composing the extracted features, the input variants and other process variants can be derived.
This way, complex artifact-centric process variants can be designed more efficiently and their quality improves, since well-known modifications are applied.
	
%Many different artifact-centric process variants, with considerable overlap.
%How to extract features from the overlap such that the features can generate the variants and possibly more models?
 
\end{abstract}

\section{Introduction}
Many business processes in modern organizations rely on knowledge workers that have to make informed decisions about specific cases. The available data and knowledge drives the decision making and processes in such knowledge-intensive processes (KiPs) \cite{CiccioM015}. 
%The structure of these knowledge-intensive business processes unpredictable and emerging, me .  KiPs are therefore non-repeatable, but there may exist parts that are repeated across several KiPs. \cite{CiccioM015}
%Case
%To support them using process automation, however, there should be repeatable parts. 
%Case management processes: unique yet repetable.
%
%
Case management is a key paradigm in BPM to support KiPs \cite{Mastering-the-Unpredictable}.
A key notion in case management is that of a \term{template}: a representation of a ``baseline'' process. The template is modified to suit the needs of the particular case being processed. Common modifications are adding and deleting elements of the template \cite{EshuisHY19}.
Such modifications result in a case management process model variant, or a \term{process variant} for short. %[What kind of modification? Subtractions allowed?]

Modifying a case management template is labor intensive, since the exact changes need to be explicitly specified. Moreover, the changes may have undesirable side effects, for instance a deadlock or a task that is done twice. 
This paper develops support for \emph{reuse of modifications} to case management templates.
By applying a modification that already was applied before to the template for another variant, the design time for new variants is reduced. Moreover, the quality of new process variants is improved, since well-proven modifications are applied.

%OVer time, a lot of variants based on the same template can be defined. Some modifications can be shared. In that case, it makes sense to distill from the variants common modifications. 
More concretely, this paper defines an approach to \emph{extract fragments from a case management template and a set of process variants that are based on the template}.  
Each fragment represents  work done for a case that was incorporated in an input process variant by modifying the template.
 We view these extracted fragments as \term{features}. %A feature is a process fragment while a feature refinement specifies how the process fragment is glued to another process variant. 
The notion of feature comes from the field of Software Product Line Engineering \cite{bookApel}, 
where they are used to distinguish common and variable parts in software artifacts and this way support reuse of software artifacts. 
Composing different but related features yields different variants of a software product \cite{ApelKL13,BatorySR04}.  
%In earlier work, we defined a feature-oriented composition approach for case management fragments \cite{Eshuis18b}.
%Also the template itself  is a feature.

%Applying a feature refinement means the original model is extended, but no behaviour is removed.

As host modeling language, we use Guard-Stage-Milestone (GSM) schemas, a technique to declaratively model life cycles of key business entities, called business artifacts \cite{Hull:GSM:10:WSFM,GSM:Info-Systems-2013}. 
In previous work, we defined a feature composition operator for GSM schemas \cite{Eshuis18b}, which defines how a GSM fragment viewed as feature is applied to a base GSM schema (template). 
Using feature composition, extracted GSM fragments can be composed into variants in a declarative way.
%By composing one or more features with the template, all the input process variants can be derived, but additional variants are possible. % \cite{Eshuis18b}. 
GSM schemas are one of the predecessors of CMMN \cite{OMG:CMMN:Beta1,Marin}.
Thus, the results in this paper provide a basis for applying feature extraction to CMMN models.
%In previous work, we developed a feature composition approach for case management processes \cite{Eshuis18b}.
%Using the discovered features reduces the design time for engineering new variants. Finally, applying well-known modifications  improves the quality of newly engineered variants.

%The approach can also be applied if one or more variants have reduced the template. In that case, the common intersection of these variants is the new template; each variant, including the old template extends this new template. 

%If a variant contains a combination of reduction and extension, we assume it can be split in two variants, one containing the reduction, one the reduction.

%
%Requirements on set of features
%- features are orthogonal, i.e., they do not overlap.
%- features are minimal, i.e., there is not another feature 
%- features are effective, i.e., they can reproduce the set of variants

The remainder of this paper is organized as follows. 
Section~\ref{sec.mot} introduces GSM schemas, feature composition and the problem of extracting features from GSM variants. Section~\ref{sec.gsm} defines GSM schemas and feature composition. Section~\ref{sec.approach} defines the approach for templates and variants that all refine the template.
%Section~\ref{sec.red} leverages the approach for GSM schemas that have been derived from the template by deleting parts of the template.
Section~\ref{sec.rel} discussed related work. Finally, Section~\ref{sec.conc} concludes the paper.

\def \BCAone {{\textsf{BCA}^1}}
\def \BCAtwo {{\textsf{BCA}^2}}
\def \BCAfour {BCA$_4$}
\def\BaseSchema{{\textsf{BCA}}^{base}}
\def\BaseSchemaRef{{\textsf{BCA}_{refactored}}^{base}}
\def \ModSchema {{\textsf{\small BCA}}^{alt\_check}}
\def \CCSchema {{\textsf{\small BCA}}^{cred\_check}}
\def \AMCSchema {{\textsf{\small BCA}}^{amc\_check}}
\def\FoneSchema{{\textsf{BCA}}^{\textit{F1}}}
\def\FoneSchemaRef{{\textsf{BCA}_{refactored}}^{\textit{F1}}}
\def\FtwoSchema{{\textsf{BCA}}^{\textit{F2}}}
\def\FtwoSchemaRef{{\textsf{BCA}_{refactored}}^{\textit{F2}}}
\def\FthreeSchema{{\textsf{BCA}}^{\textit{F3}}}
%
%%To maintain the processes, 

%\vspace*{-2mm}

\section{Overview}\label{sec.mot}

We introduce GSM schemas, feature-based composition of GSM fragments, and the problem of extracting features from a set of   GSM schema variants.

\paragraph{GSM schemas.}
A Guard-Stage-Milestone (GSM) schema defines the life cycle of a business artifact \cite{GSM:Info-Systems-2013,Hull:GSM:10:WSFM}. A business artifact is a key business entity that is changed during a business process, for instance an order or a claim request \cite{Nigam2003}. Key modeling constructs in GSM schemas are stages and milestones. A stage represents a cluster of business activity performed for the artifact. Stages are organized into a hierarchy. Each atomic stage contains exactly one task, an atomic piece of work. A milestone represents a business objective, usually achieved by completing an attached stage. Stages and milestones change status if certain conditions, called sentries, are met. There are two kinds of sentries. Plus sentries ensure that a stage is opened or a milestone achieved, while minus sentries ensure that a stage is closed or milestone invalidated. Guards are plus sentries of stages.

\begin{figure}[t]\centering
	\includegraphics[scale=0.5]{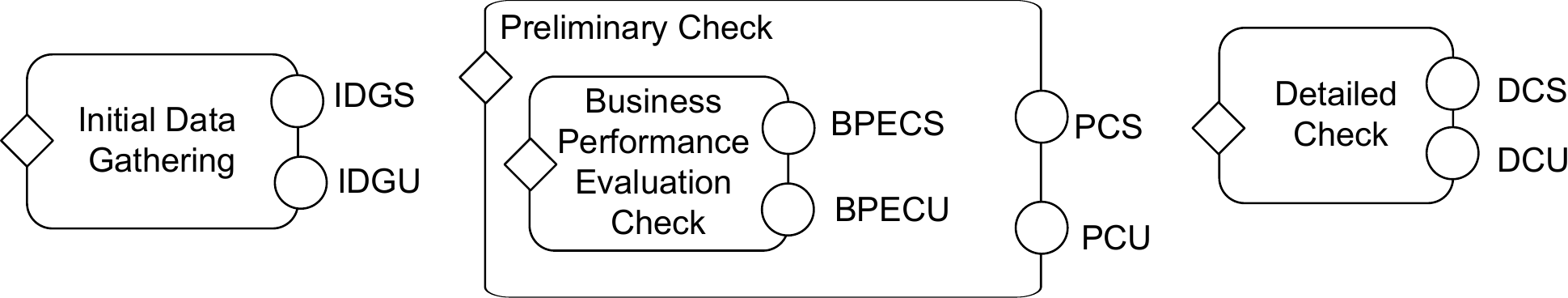}
	\caption{\label{fig:ex1} Base GSM schema Business Criteria Assessment ($\BaseSchema$)}
\end{figure}

%
%\begin{figure}[tb]
%	\centerline{
%		%        \begin{tabular}{cc}
%		%        \begin{subfigure}[b]{1.0\textwidth}
%		\includegraphics[scale=0.5]{ex1-hierarchy}}
%	\caption{\label{fig:ex1} Base GSM schema Business Criteria Assessment ($\BaseSchema$)}
%\end{figure}

%\paragraph{Feature-oriented composition.}

%A feature is a specific functionality of a software artifact that is discernible for an end user \cite{bookApel}. In this paper, each feature denotes a GSM schema fragment.

\begin{figure}[t]\centering

%\subfloat[\label{fig:ex1} Base GSM schema Business Criteria Assessment ($\BaseSchema$)	]{	\includegraphics[scale=0.5]{ex1-hierarchy}}

%\begin{subfigure}[tb]
%	\centerline{
		%        \begin{tabular}{cc}
		%        \begin{subfigure}[b]{1.0\textwidth}
\subfloat[\label{fig:ex3} Variant $\BCAone$ of $\BaseSchema$]{\includegraphics[scale=0.5]{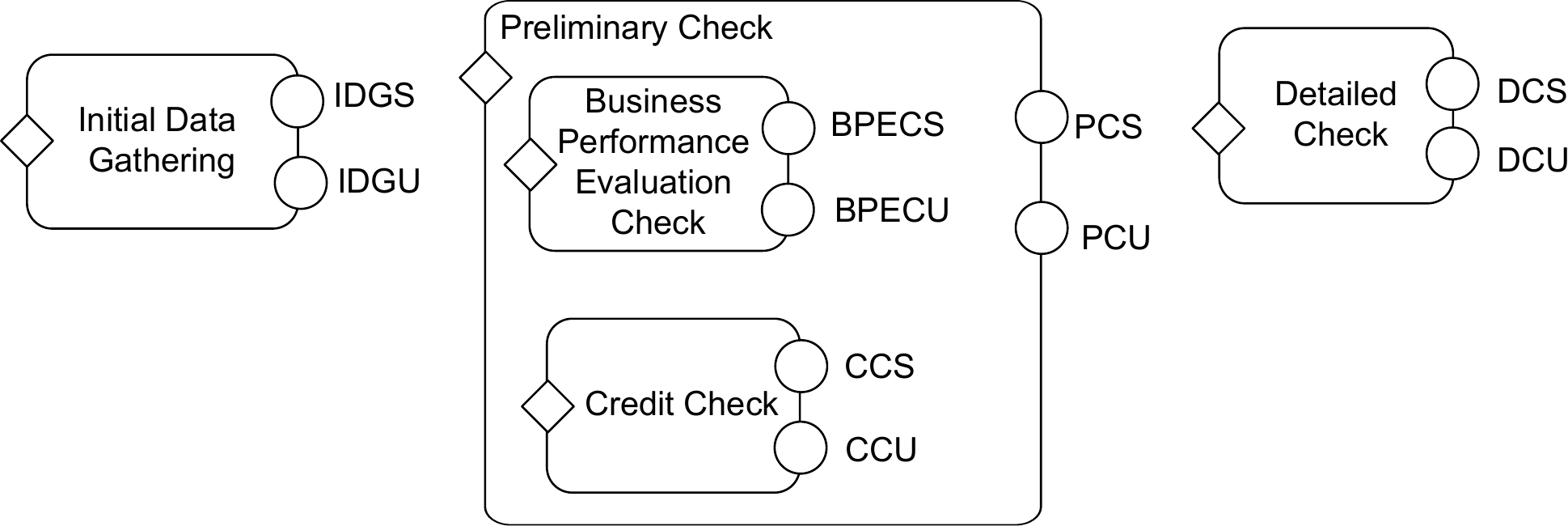}}
%		}
%	\caption{}
%\end{subfigure}

%\begin{subfigure}[tb]
%	\centerline{
		%        \begin{tabular}{cc}
		%        \begin{subfigure}[b]{1.0\textwidth}
\subfloat[\label{fig:ex4} Variant $\BCAtwo$ of $\BaseSchema$]{\includegraphics[scale=0.5]{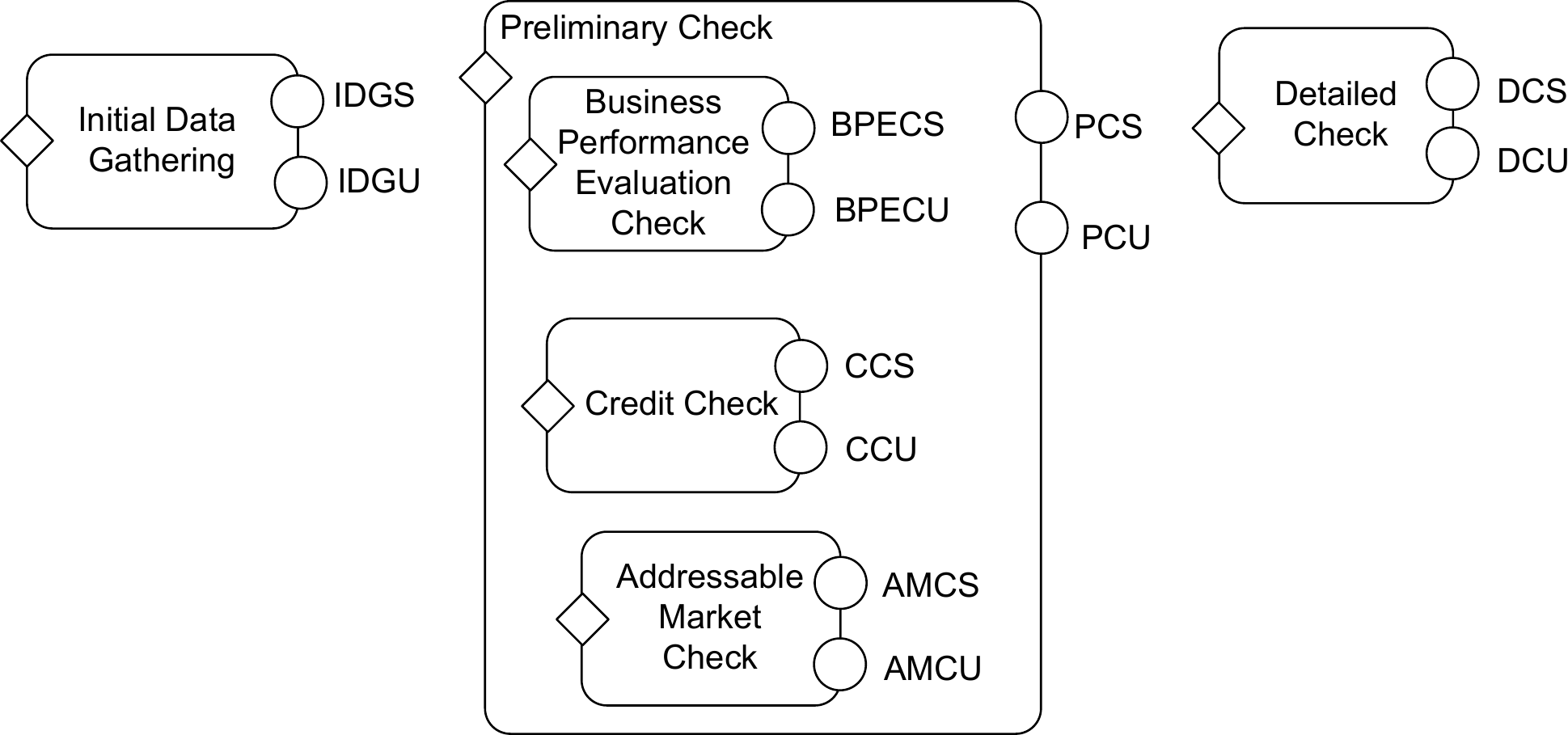}}
%	}
%	\caption{\label{fig:ex4} Variant  $\BCAtwo$ of $\BaseSchema$}
%\end{subfigure}
%

%\begin{subfigure}[tb]
%	\centerline{
		%        \begin{tabular}{cc}
		%        \begin{subfigure}[b]{1.0\textwidth}
%\subfloat[\label{fig:ex5} Variant  $\BCAthree$ of $\BaseSchema$]{\includegraphics[scale=0.5]{ex5-hierarchy}}
	%	}
%	\caption{\label{fig:ex5} Variant  $\BCAthree$ of $\BaseSchema$}
%\end{subfigure}
\caption{\label{fig1} Variant GSM schemas}
\end{figure}
\paragraph{Example.}
Fig.~\ref{fig1} shows sample GSM schemas. Rounded rectangles denote stages, circles denote milestones, and each diamond denotes the presence of a guard. Other sentries are not visualized.  In the base process (Fig.~\ref{fig:ex1}), business criteria for a partner contract are assessed: first data about the partner is gathered and prechecked, and next a detailed check is performed to decide whether the criteria should be changed or not. If new information arrives before the business criteria have been assessed, the data is gathered anew and the business criteria check is restarted, if applicable. Sentries for the base process are listed in Table \ref{tab.stg} and \ref{tab.mst}.

\begin{table}[tb]{
	\caption{Stages and sentries  for $\BaseSchema$ in Fig.~\ref{fig1}\label{tab.stg}. `;' separates different sentries}
			\centerline{\footnotesize
			\begin{tabular}{p{0.25\linewidth}p{0.25\linewidth}p{0.42\linewidth}}
				\toprule
				{\bf Stage} & \bf Plus sentries (guards) & \bf Minus sentries (closing) \\ \midrule
				{\sf Initial Data Gathering} & \sf E:StartAssessment\,\textnormal{;} \sf E:AdditionalInfo &{\sf IDGS}  \\
				{\sf Preliminary Check} & {\sf IDGS} & \sf PCS $;$ PCU\,\textnormal{;} E:AdditionalInfo\\
				{\sf Business Performance Evaluation Check} & {\sf  +Preliminary Check}  & \sf BPECS $;$ BPECU\,\textnormal{;} \sf -Preliminary Check\\
				% \sf Addressable Market Check & \sf IGDS $\land annual\_revenue \geq \$500K$ \\
				\sf Detailed Check & \sf PCS & \sf DCS $;$ DCU  \\ \bottomrule
%				\sf Fast Turnaround Bonus Eligibility  & \sf IDGS \\
%				\sf Team Bonus Pay  & \sf C{:}Fast Turnaround Bonus Eligibility %\\                     & \ \ \ $\land$ \sf fast\_turnaround 			$\land$ PCS \\
	\end{tabular}}}
\end{table}

\begin{table}[tb]
	\caption{Milestones and sentries for $\BaseSchema$ in Fig.~\ref{fig1}\label{tab.mst}}
			\centerline{\footnotesize
			\begin{tabular}{p{0.09\linewidth}p{0.37\linewidth}p{0.335\linewidth}p{0.15\linewidth}}
				\toprule
				\bf Mile\-stone & \bf Full Name & \bf Plus sentries  & \bf Minus sentries   \\ \midrule
				\sf IDGS & \sf Initial Data Gathering Successful & \sf C:Initial Data Gathering & \sf E:AdditionalInfo \\
	%			\sf IDGU & \sf Initial Data Gathering Unsuccessful & \sf C:Initial Data Gathering $\land \ldots$& \sf false\\
%				\sf CCS & \sf Credit Check Successful & \sf \sf C:Credit Check $\land$ rating $\geq$ 8 \\
%				\sf CCU & \sf Credit Check Unsuccessful & \sf \sf C:Credit Check $\land$ rating $< 8$  \\
				\sf BPECS & \sf Business Performance Evaluation Check Successful &  \sf C:Business Performance Evaluation Check $\land$ \sf BP\_good& \sf \sf E:AdditionalInfo\\
				%       & \ \ \ \sf   & \ \ \ \sf  \\
				\sf BPECU & \sf Business Performance Evaluation  Check Unsuccessful& \sf C:Business Performance Evaluation  Check $\land \ \neg$\sf BP\_good& \sf \sf E:AdditionalInfo \\
				%      & \ \ \ \sf  & \ \ \ \sf  \\
				%& \sf AMCS & \sf Addressable Market Check Successful & \sf C:Addressable Market Check $\land \ldots$\\
				%& \sf AMCU & \sf Addressable Market Check Unsuccessful  & \sf C:Addressable Market Check $\land \ldots$ \\
				%& \sf Addressable Market Check & \sf IGDS $\land annual\_revenue \geq \$500K$ \\
				\sf PCS & \sf Pre-checks Successful & \sf BPECS & \sf false \\
				\sf PCU & \sf Pre-checks Unsuccessful & \sf BPECU  & \sf false\\
				\sf DCS & \sf  Detailed Check Successful & \sf C:Detailed Check $\land \ \dots$& \sf false \\
				\sf DCU & \sf  Detailed Check Unsuccessful & \sf C:Detailed Check $\land \ \dots$& \sf false \\ \bottomrule
%				\sf TBPS & \sf Team Bonus Pay Successful & \sf C:Team Bonus Pay  \\
	\end{tabular}}
\end{table}

\paragraph{GSM variants and feature composition.}
A GSM schema can be modified into another GSM schema. This way, from a base GSM schema a set of variant GSM schemas can be derived. 
To derive a variant, several change operations can be applied to a GSM schema \cite{EshuisHY19}.
For instance, Fig.~\ref{fig1}\subref{fig:ex3} and Fig.~\ref{fig1}\subref{fig:ex4} show two variants that are modifications of the base GSM schema in Fig.~\ref{fig:ex1}.
They have been derived by inserting  stages and milestones for an additional checks on the credit level and the addressable market. 
%The last variant has been derived by both inserting stages and attached milestones and deleting stage \textsf{Detailed Check} and its attached milestones. 
Also sentries need to be modified for these variants: Table \ref{tab.mst-var-pcs} shows as an example the modified sentries for \textsf{PCS} and \textsf{PCU}. Space limitations prevent a completing listing of the  sentries for the other stages and milestones of the variants.
%: for instance, the sentry for \textsf{PCS} in $\BaseSchema$ is $\textsf{BPECS}$ while the sentry for \textsf{PCS} in $\BCAone$ is $\textsf{BPECS} \land \textsf{CCS}$, to ensure that the results of all checks are successful.

%Problem: how to identify features?

To enable reuse of shared parts between GSM variants, we proposed in earlier work  \cite{Eshuis18b} to view GSM fragments as features and to use feature composition, well-known from the field of Software Product Line Engineering \cite{ApelKL13,BatorySR04}, to compose GSM fragments into GSM schemas.
%In earlier work, we introduced feature composition for GSM schema fragments. 

\begin{table}[tb]{
		\caption{Sentries for milestones \textsf{PCS} and \textsf{PCU} in variants of $\BaseSchema$ in Fig.~\ref{fig1}.\label{tab.mst-var-pcs}}
		\centerline{\footnotesize
			\begin{tabular}{p{0.23\linewidth}p{0.1\linewidth}p{0.425\linewidth}p{0.2\linewidth}}
				\toprule
				\bf Variant &	\bf Mile\-stone  & \bf Plus sentries (achieving) & \bf Minus sentries (invalidating)  \\ \midrule
				$\BCAone$ &			\sf PCS  &  \sf BPECS $\land$ CCS  & \sf false \\
				$\BCAone$ &			\sf PCU  & \sf BPECU $;$ CCU  & \sf false \\
				$\BCAtwo$ &			\sf PCS  & \sf BPECS $\land$ CCS $\land$ AMCS  & \sf false \\
				$\BCAtwo$ &			\sf PCU  & \sf BPECU $;$ CCU $;$ AMCU  & \sf false \\ \bottomrule			
	\end{tabular}}}
\end{table}

\paragraph{Extracting features.}

Though the feature composition approach \cite{Eshuis18b} allows to generate different variants from a common base GSM schema (template), it  assumes that the composed features  already exist. %This means that features need to be defined upfront. 
But defining features manually can be time consuming and costly, as it depends on domain knowledge from domain experts.

%Though GSM schema variants can be different, certain parts may have been derived in a similar way from a base GSM schema. 
%To allow for reuse of modifications, 
We define an approach to extract features from a set of variant GSM schemas derived from the same base GSM schema.
The approach is  efficient, since it can be automated and does not rely on domain knowledge, because that knowledge is already encoded in the different variants. 
Moreover, extracting features enables the reuse of   modifications among different GSM variants, each feature representing one set of related modifications.

In essence, the approach decomposes a base GSM schema and set of GSM schema variants into features.
For example, for the base schema $\BaseSchema$ and its two variants $\BCAone$ and $\BCAtwo$, two features can be extracted, one related to Credit Checking, one  related to Addressable Market Check. Each feature represents the insertion of a stage and two milestones and the modification of sentries.
Each input GSM schema variant can be derived by composing one or more of the extracted features with the base schema, using feature composition \cite{Eshuis18b}. 
However, also other variants can be derived by composing the extracted features. 
For instance, another variant for $\BaseSchema$ can be composed that contains \textsf{Addressable Market Check} and its connected milestones, but not \textsf{Credit Check}.

\section{ GSM Schemas}\label{sec.gsm}
\label{sec:GSM-formal}
%\vspace*{-3mm}

%We define the syntax and  semantics of GSM schemas in this section.

In this section, we define GSM schemas.
A GSM schema~\cite{GSM:Info-Systems-2013}  of a business artifact  consists of data attributes and status attributes. 
Data attributes model the information state of the business artifact.
A status attribute is a Boolean variable that denotes the status of a stage
or milestone. 
For a status attribute of a stage, value \textit{true} denotes that the stage is open, value \textit{false} that the stage is closed.
For a status attribute of a milestone, value \textit{true} denotes that the milestone is achieved, value \textit{false} that the milestone is invalid.

Event-Condition-Action rules define for which event under which condition a status attribute changes value (action). 
The event-condition part of a rule is called a \emph{sentry}. 
The event of a sentry  is optional. 
We distinguish between external and internal events. An external event signifies a change in the environment. It is either a task completion event $\cmpltn{T}$, where $T$ is a task, as defined below, or a named external event $\evt{n}$, where $n$ is an event name. An internal event signifies a change in value of a status attribute $a$: internal event $+a$ denotes that  $a$ becomes {true}, $-a$ that $a$ becomes {false}. 
For instance, \textsf{\sf +Preliminary Check} in Table~\ref{tab.stg} is an internal event that signifies that stage \textsf{Preliminary Check} gets opened.
%For instance, in condition $\textsf{\small IDGS} \land \textsf{\small employee\_count} \geq 300$ (Table~\ref{tab.fr1}),  \textsf{\small IDGS} is a milestone while \textsf{\small employee\_count} is a data attribute.
The condition of a sentry is a Boolean expression that can refer to data attributes or status attributes. 
%Note that an internal event $+a$ is only true during a B-step. %, while a condition $a$ can be true indefinitely long. 
%
The action of each rule is that a status attribute becomes {true} or {false}, which leads to two types of sentries. 
%Given these two distinct actions, 
%Consequently, 
%we distinguish between two types of sentries. 
A \emph{plus sentry} defines when a stage becomes open or a milestone gets achieved. A \emph{minus sentry} defines when a stage is closed or a milestone gets invalid. 
%For each plus (minus) sentry, there is an associated plus (minus) rule.
%For instance, the plus sentry of stage \textsf{\small Business Performance Evaluation Check}  is \textsf{\small \sf +Preliminary Check}. 

Stages and milestones can be nested inside other stages. A milestone cannot contain any other milestone or stage. We require that the nesting relation induces a forest, i.e., the nesting relation is acyclic and if a stage or milestone is nested in two other stages $S_1$, $S_2$, then either $S_1$ is nested in $S_2$ or $S_2$ in $S_1$. 
%A task cannot contain itself a stage or milestone.
The most nested stages, which are called \term{atomic},  launch tasks. 
To ease the presentation, we assume for this paper that stages  launch tasks having the same label, so for instance stage \textsf{Detailed Check} launches a task with the same name. 
%
%TaskStage completion events only exist for the most nested stages, which are called \emph{atomic}.

We next formally define GSM schemas \cite{GSM:Info-Systems-2013,EshuisHY19}. 
%We assume a global universe $\caU$ of named external events and attributes, partitioned into sets of named external events $\Ue$, data attributes $\Ud$, stage attributes $\US$, and milestone attributes $\UM$. 

\begin{deff}[GSM schema]
%\label{def:GSM}
%\vspace*{-1mm}
A  {\em GSM schema} is a
tuple $\Gamma=(\Att=\Data  \cup \Stg \cup \Mst,\Ev=\EvE\cup \EvC,\preceq,\rules=\rules_+ \cup \rules_-)$, where
\begin{itemize} %[$\LargerCdot$]
	\item $\Att$ is a set of attributes, partitioned by the following three subsets:
\begin{itemize}
	\item $\Data$ is a finite set of data attributes;
	%\item $\Task $ is a finite set of task attributes;
	\item $\Stg $ is a finite set of stage attributes;
	\item $\Mst $ is a finite set of milestone attributes;
\end{itemize} 
%\item $\Task$ is a finite set of tasks;
\item $\EvE=\{~ \evt{n} ~|~ n \textnormal{ is an event name}~\}$ is a finite set of named external events;
\item $\EvC=\{ ~\cmpltn{S} ~|~ S \in \Stg_{atomic}~\}$ is the set of stage completion events;
\item $\preceq \mathop{\subseteq} ( \Stg \cup \Mst) {\times} \Stg$ is a partial order on stages and milestones, where $a_1 \preceq a_2$ means that $a_1$ is child of $a_2$. Relation $\preceq$  induces a forest, i.e., if $a_1 \preceq a_2$ and $a_1 \preceq a_3$, then $a_2 \preceq a_3$ or $a_3 \preceq a_2$. We let $\Stg_{atomic}$ denote the set of stages that have no children;
%\item $\task: \Stg_{atomic} \fun \Task$ is an injection from atomic stages to tasks;
 \item $\rulesPlus$, $\rulesMinus$ are functions assigning to each status attribute $ \Stg \cup \Mst$ non-empty sets of \term{sentries} (see Definition~\ref{def2}). For $ a\in  \Stg \cup \Mst$, $\rulesPlus(a)$ is the set of {\em plus sentries}  that define the conditions when to open  stage $a \in  \Stg $ or achieve milestone $a\in \Mst$, while $\rulesMinus(a)$ is the set of {\em minus sentries}  that define the conditions when to close stage $a\in  \Stg$ or invalidate milestone $a\in \Mst$. 
\end{itemize}
\end{deff}
%Stage $S$ is atomic if there is no other stage $S'\in \Att_{S}$ such that $S'\preceq S$. The definition of~$\preceq$ ensures that tasks and milestones are atomic by default.
Relation $\preceq$ is visualized using nesting. For instance,  \textsf{Business Performance Evaluation Check} $\preceq$ \textsf{Preliminary Check} and \textsf{BPECS} $\preceq$ \textsf{Preliminary Check} in Fig.~\ref{fig1}\subref{fig:ex1}. 

Each sentry $\varphi$ in set $\rulesPlus(a)$, where $a \in \Stg\cup\Mst$, maps into an Event-Condition-Action rule ``$\varphi\ \THEN\ {+a}$'', where sentry $\varphi$ is the Event-Condition part and action $+a$ denotes for $a\in \Stg$ that stage $a$ gets opened and for $a\in \Mst$ that milestone $a$ gets achieved.
%, if $a\in \Mst$. 
Each sentry $\varphi$ in set $\rulesMinus(a)$ maps into a rule ``$\varphi \ \THEN\ {-a}$'', where action $-a$ denotes for  $a\in \Stg$ that stage $a$ gets closed  and  for $a \in \Mst$ that milestone $a$ gets  invalid. 
%For example, the plus sentry  for milestone \textsf{BPECS}  (cf.\ Table~\ref{tab.mst}) is \cmpltn{\textsf{Business Performance Evaluation Check Successful}} $\land \ \textsf{BP\_good}$; the corresponding Event-Condition-Action rule is \cmpltn{\textsf{Business Performance Evaluation Check Successful}}$\ \land\ \textsf{BP\_good} \ \THEN\ {+}\textsf{BPECS}$.
Each sentry in set $\rulesPlus(a)$ or $\rulesMinus(a)$  is sufficient for triggering a status change in the stage or milestone $a$. 

For the definition of sentries, we assume a condition language $\caC$ that includes predicates over integers and Boolean connectives. The condition formulas may refer to stage, milestone and data attributes from the universe of attributes $\caU$. Keyword $\orig$ denotes the original condition formula defined in another GSM schema for the same status attribute \cite{Eshuis18b}.

\begin{deff}[Sentry]\label{def2}
	A  {\em sentry} has the form $\tau \land \gamma$, where $\tau$ is the event-part and $\gamma$ the condition-part. The event-part $\tau$ is either empty (trivially true), a named external event $E$, a task completion event $\cmpltn{T}$, where $T$ is a task, or is an internal event $+a$ or $-a$, where $a $ is a stage or milestone attribute. 
	The condition $\gamma$ is a Boolean formula in CNF in the condition language $\caC$ that refers to $\Att \cup \{\orig\}$, so  data attributes in $\Data$ and status attributes in $\Stg \cup \Mst$ and the keyword $\orig$ can be used in $\gamma$. The condition-part can be omitted if it is equivalent to $true$. %To define feature composition, we use $\orig$ as special keyword in $\caC$.
\end{deff}
%For instance, the plus sentry of stage \textsf{Business Performance Evaluation Check} (cf.\ Table~\ref{tab.stg}) references  internal event \textsf{+Preliminary Check} (generated when the corresponding stage gets opened) and the condition-part is omitted. 
%%Internal event \textsf{+Preliminary Check} is generated when stage  \textsf{Preliminary Check} gets opened. 
%The plus sentry of stage \textsf{Detailed Check} has an empty event-part and condition \textsf{PCS}, which is true if milestone \textsf{PCS} has been  achieved.
%

%Note that a sentry  in a GSM schema $\Gamma$ refers to attributes in $\caU$ that may or may not be defined in $\Att$. We next define a subclass of GSM schemas that are self-contained, i.e., schemas whose rules only reference attributes that are are defined in the schemas themselves.

%Sentries containing $\orig$ refer to other sentries defined in other GSM schemas. We next define a subclass of GSM schemas that are self-contained.
The condition part of a sentry is a boolean formula in conjunctive normal form (CNF) in order to ease the presentation. However, this is not a severe restriction: if the condition part of a formula in $\caC$ is not in CNF, it can be rewritten using Boolean laws into an equivalent set (disjunction) of sentries. For instance,  formula $\evt{\textsf{n}} \land (a>10 \lor b <5) $ is equivalent to set of sentries $\{\evt{\textsf{x}} \land a>10, \evt{\textsf{x}} \land b<5\}$.

In earlier work we proposed a declarative composition operator $\after$ for GSM schemas \cite{Eshuis18b}. 
Given a GSM schema fragment $\GammaOne$ and a GSM schema $\GammaTwo$, the GSM schema $\GammaOne \after \GammaTwo$ results by merging the GSM schemas by taking the union of the different components of a GSM schema tuple, except for sentries (rules). For shared stages and milestones the definition of sentries in $\GammaOne$ override those in $\GammaTwo$. 

If the sentries of a shared stage or milestone should be merged rather than overridden, keyword $\orig$ can be used in the sentries of $\GammaOne$, which in the evaluation of $\after$ is replaced with the sentries of the shared stage or milestone in $\GammaTwo$. For instance, if $\GammaOne$ and $\GammaTwo$ share milestone $m$, the sentry of  $m$ in $\GammaTwo$ is $x >10$  and in $\GammaTwo$ is $\orig \land x<100$ , then the sentry of  $m$ in $\GammaOne \after \GammaTwo$ is $x >10 \land x <100$.

GSM schema fragments contain $\orig$ in their sentries, while base schemas and  variants do not contain sentries with $\orig$, i.e., they can be executed.

\def\eDG{\textit{erDG}}

\section{Extracting Features}\label{sec.approach}

We next define a method that extracts a set of features from a template (base) GSM schema and a set of GSM schema variants that refine the template.
Each extracted feature is specified as a GSM schema fragment.
%Each  variant can be derived by composing the template with one or more of the extracted features. 

\subsection{Requirements} 
First, we list  requirements on the set of features that the method extracts:
%Requirements
\begin{itemize}
    \item[R1] The feature set must be \emph{minimal}: there is no feature in the set that itself is a combination of other features in the set.
    \item[R2] The feature set must be \emph{complete}: each variant can be derived by composing  one or more features with the template.
    %\item  <The features must be orthogonal: there are no two features in the set such that they overlap. What about overlapping features?]
  %   \item The features are permutable.
\end{itemize}
Requirement R1 states that features are not overlapping, so orthogonal. For instance, variant $\BCAtwo$ can be derived by applying two features, one of which is defined for generating variant $\BCAone$. Introducing a third feature that derives $\BCAtwo$ directly from $\BaseSchema$ is therefore redundant.
Requirement R2 ensures that each variant that is input can in fact be derived by a combination of features.

\subsection{Method}
\begin{figure}[t]
\centerline{\includegraphics[scale=0.55]{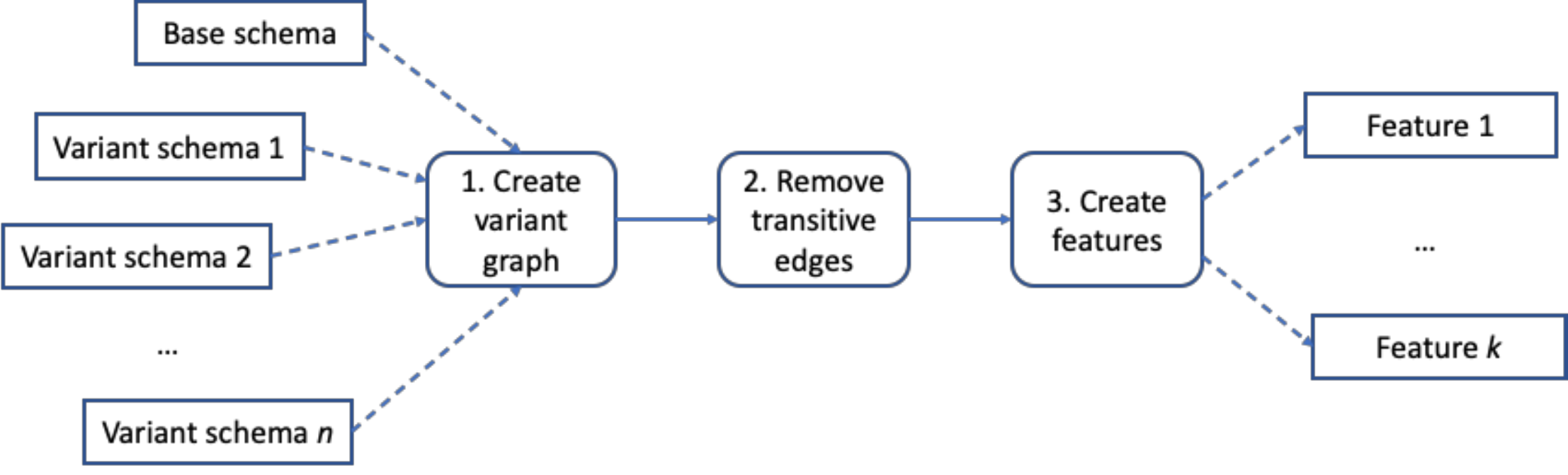}}

%%\begin{algorithm}[t]
%\begin{algorithm}[t]
%\begin{algorithmic}[1]
%	\Require  Template GSM schema $\base$ and set of GSM schemas $\{\Gamma^1,\upto,\Gamma^n\}$ 
%	such that each refines $\base$
%	\Ensure Set of features $F$ such that all GSM schemas in $\{\Gamma^1,\upto,\Gamma^n\}$ can be generated by composing one or more features in $F$ with $\base$ 
%%	\item 1. GSM core is intersection
%\Statex
%\State Create a variant graph. Nodes are $\base$ and the GSM schemas in $\{\Gamma^1,\upto,\Gamma^n\}$. An edge exists from GSM schema $\Gamma$ to $\Gamma'$ if $\Gamma'$ refines $\Gamma$, written $\Gamma \subseteq \Gamma'$.
%    \State  Perform transitive reduction: For each edge $(\Gamma,\Gamma')$ from the graph, remove the edge if there is an alternative path from $\Gamma$ to $\Gamma'$.
%    \State Let $F=\emptyset$. For each edge $(\Gamma,\Gamma')$, create a feature $f=\Gamma'\setminus \Gamma$ and add it to $F$.
%    \State \textbf{return} $F$ 
%   % \item  Remove duplicate feature [ redundant if feature set]
%% 	\item 2. GSM feature $i$ is GSM $i$ minus GSM base. GSM feature either using merge orig or override
%	%	\item 2. GSM feature 2 is GSM two minus GSM core
%\end{algorithmic}
\caption{Method to extract features from a set of GSM schemas that all refine the same base GSM schema \label{fig.method}}
%\end{algorithm}
\end{figure}

%Next, we define the method. 
The input, output and different steps of the method are shown in Fig.~\ref{fig.method}.
We next explain and define the different steps in detail.

\def \base {\Gamma^{base}}

\paragraph{Step 1}
%\begin{wrapfigure}{r}{.4\textwidth}
	\begin{figure}[t]
\centerline{\includegraphics[scale=0.5]{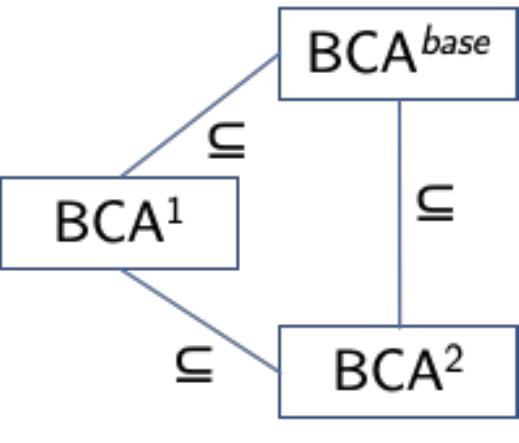}}
\caption{\label{fig.BCAref}Variant graph for GSM schemas  in Fig.~\ref{fig:ex1} and~\ref{fig1}}
%\end{wrapfigure}
\end{figure}
 creates 
a variant graph that shows the refinement relations between the different GSM schema variants. Nodes are $\base$ and the variant GSM schemas. An edge $(\Gamma,\Gamma')$ in this variant graph denotes that $\Gamma'$  refines (extends)  $\Gamma$, written $\Gamma \subseteq \Gamma'$.  Fig.~\ref{fig.BCAref} shows the variant graph for the GSM schema variants in Fig.~\ref{fig1}, which all  refine $\BaseSchema$.  Moreover, $\BCAtwo$ refines $\BCAone$.

The refines relation, $\subseteq$, is defined as follows. 
%First we define GSM schema containment.
\begin{deff}[Refinement]
	Let $\GammaOne, \GammaTwo$ be two GSM schemas.
	Then $\GammaTwo$ refines $\GammaOne$, written $\GammaOne \subseteq \GammaTwo$, if 
	\begin{itemize}
		\item $\Data^{1} \subseteq \Data^{2}$;
		\item 
		$\Stg^{1} \subseteq \Stg^{2}$;
		\item 
		$\Mst^{1} \subseteq \Mst^{2}$;
	%	\item 
	%	$\Task^{1} \subseteq \Task^{2}$;
		\item $\EvE^{1} \subseteq \EvE^{2}$;
		\item 
		$\EvC^{1} \subseteq \EvC^{2}$;
		\item $ \preceq^{1} \subseteq \preceq^{2}$;
%		\item $\tau^1 \subseteq \tau^2$;
		%\item $\rulesPlus=\rulesPlus^{1} \cup \rulesPlus^{2}$;
		\item for each $a \in \Att^1$, if $\varphi \in \rulesPlus^1(a)$ then  there is a  $\varphi' \in \rulesPlus^2(a)$ s.t.\ $\varphi$ implies $\varphi'$;
		%, in other words $\varphi=\varphi'$ or $\varphi$ is a conjunct of $\varphi'$ [what if $\varphi$ is false?]
				\item for each $a \in \Att^1$, if $\varphi \in \rulesMinus^1(a)$ then  there is a $\varphi' \in \rulesMinus^2(a)$ s.t.\ $\varphi$ implies $\varphi'$.
%		\item for each $a \in \Att^2$, if $\varphi \in \rulesPlus^2(a)$ then  there is a $\varphi' \in \rulesPlus^1(a)$ such that $\varphi$ is a conjunct of $\varphi'$
		
	\end{itemize}
	
\end{deff} 
In the definition, most lines are straightforward given the definition of GSM schemas, except the one about the rules. If $\GammaOne \subseteq \GammaTwo$ then for each plus (minus) rule $\varphi$ of status attribute $a$ in $\GammaOne$, the same plus (minus) rule  or plus (minus) rule  $\varphi \land \psi$ for $a$ is in $\GammaTwo$. If the rule is of the form $\varphi \land \psi$ in $\GammaTwo$, then a feature can be constructed that extends $\varphi$ with conjunct $\psi$, by defining a sentry $\orig \land \psi$. A rule in $\GammaTwo$ of the form $\varphi \lor \psi$ is not allowed (cf.\ Def.\ \ref{def2}); instead, a set of  rules $\{\varphi,\psi\}$ can be specified.

%on E if a<10
% on E if a < 10 or b>5

\paragraph{Step 2} removes each transitive edge from the variant graph, i.e., each edge $(\Gamma,\Gamma')$  for which there is an alternative path from $\Gamma$ to $\Gamma'$. Step 2 is needed to ensure that the feature set is minimal. A refinement relation that is implied by one or more other refinement relations can be safely deleted, since its effect can be obtained from the  other refinement relations. 
For the variant graph in Fig.~\ref{fig.BCAref}, edge $(\BaseSchema,\BCAtwo)$ is transitive and removed in step 2.
%For each extension relation a feature is created in Step 3. 
%By removing refinement relations from the graph that are already implied by the other refinement relations, the method  creates a minimal set of features.  

%What if there are alternative, disjoint paths from $g$ to $h$? Permutability? Yes, if two features refine the same model elements, then they should do so by either extending or merging.

%For instance, $CCS and BPECS$ vs $BPECS ; employee\_ count <300 and IDGS$. These are two different variants. $CCS AND employee\_count <300 and IDGS$ and $employee\_count <300 and IDGS$. Only one final variant is allowed, so the feature dependency can be derived from the final variant.

\paragraph{Step 3} creates 
for each edge $(\Gamma,\Gamma')$ in the variant graph a feature. 
Since $\Gamma \subseteq \Gamma'$, constructing a feature seems straightforward: simply delete $\Gamma$ from $\Gamma'$. 
However, to ensure that composing the feature with $\Gamma$ yields $\Gamma'$, the feature must include some attributes from $\Gamma$. % to ensure that the feature correctly refines $\Gamma$. 

%Note that the role of the template may also be performed by another process variant that is refined by the variant. In the remainder of this section, whenever the template is referred to that is refined by a process variant, also another process variant can be   

\begin{figure}[ptb]
	\centerline{
		%        \begin{tabular}{cc}
		%        \begin{subfigure}[b]{1.0\textwidth}
		(a)\includegraphics[scale=0.5]{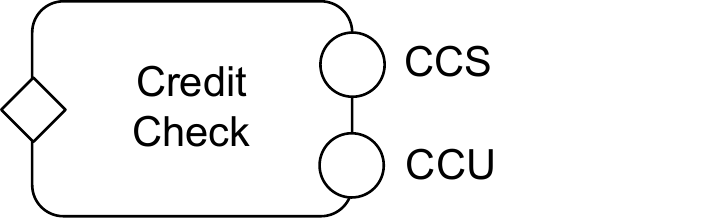}
			(b)\includegraphics[scale=0.5]{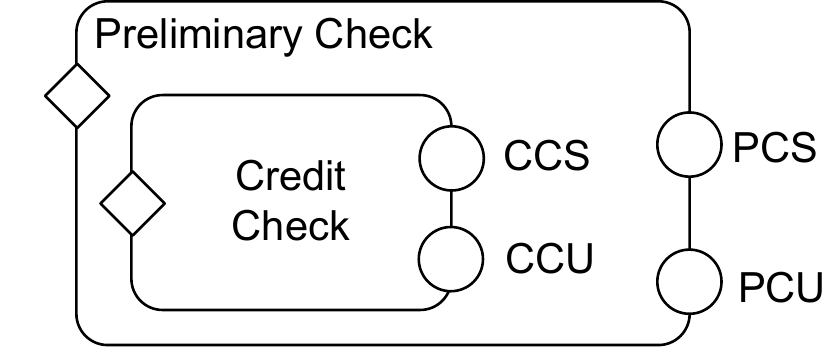}
		}
	\caption{\label{fig:ex3-feature} (a) Difference between $\BaseSchema$ and $\BCAone$; (b) Feature extracted for $\BaseSchema$ and $\BCAone$ }
\end{figure}

For instance, consider $\BaseSchema$ and $\BCAone$ in Fig.~\ref{fig:ex1} and~\ref{fig1}\subref{fig:ex3}. Deleting $\BaseSchema$ from $\BCAone$ gives the GSM schema fragment in Fig.~\ref{fig:ex3-feature}(a).  But applying this as feature to $\BaseSchema$ results in a GSM schema variant in which the result of  stage \textsf{Credit Check}, represented by milestones \textsf{CCS} and \textsf{CCU}, is not linked to the sentries of the stages and milestones of $\BaseSchema$. 
Fig.~\ref{fig:ex3-feature}(b) shows the correct feature: the fragment includes milestones \textsf{PCS} and \textsf{PCU}; the sentry of \textsf{PCS} is $\orig \land \textsf{CCS}$ while the sentries for \textsf{PCU} are $\textsf{CCU}$ and $\orig$. These sentries connect \textsf{CCS} and \textsf{CCU} to $\BaseSchema$.

Given a core feature extracted  by subtracting $\Gamma$ from $\Gamma'$, the next definition characterizes which status attributes from $\Gamma$ should be added to the core  feature, in order to derive $\Gamma'$.  We call these the border attributes, since they are the attributes from $\Gamma$ that link with the core feature.

\begin{deff}[Border attributes]\label{def.border}
	Let $\GammaOne, \GammaTwo$ be two GSM schemas such that $\GammaOne \subseteq \GammaTwo$ (e.g., $\GammaOne$ is the template, $\GammaTwo$ the variant).	%Then $borderAtts(\GammaOne \setminus \GammaTwo)$ is the set $\{a \in \Att^2|  vars(\rules(a)) \cap (\Att^1 \setminus \Att^2) \neq\emptyset \} \cup \{a \in \Att^2 | \exists b \in \Att^1 \setminus \Att^2: a \preceq^1 b \lor b \preceq^1 a \}$
	Then $borderAtts(\GammaOne, \GammaTwo)$ is the set of status attributes of $\GammaOne$ that indirectly  reference attributes in the part of  $\GammaTwo$ that is not in $\GammaOne$:
	\begin{eqnarray*}
		borderAtts(\GammaOne, \GammaTwo)  &= &
		\{a \in \Att^1|  \exists \varphi \in \rules^{1}(a): atoms(\varphi)\cap (\Att^2 \setminus \Att^{1})  \neq\emptyset \} \\
		&\cup& \{a \in \Att^1 | \exists b \in \Att^2 \setminus \Att^{1}: a \preceq^1 b \lor b \preceq^1 a \}.
	\end{eqnarray*}
\end{deff}
The definition shows that border attributes need to be included for two reasons.
First, if there is a sentry for a stage or milestone, such that the sentry references attributes of $\GammaTwo$ that are not in $\GammaOne$. 
For instance, the sentry of \textsf{PCS} of $\BCAone$ is $\textsf{BPECS} \land \textsf{CCS}$ and \textsf{CCS} is in $\BCAone$ but not in $\BaseSchema$. Therefore, \textsf{PCS} is a border attribute for $\BaseSchema$ and $\BCAone$, to ensure that the extracted feature  modifies the sentry of \textsf{PCS}.

%First, consider A B and A C B, where guard of B is "if C". The feature is C. Then B should be included in the feature to properly glue the feature to the original variant.
Second, if a stage or a  milestone that is in $\GammaTwo$ but not in $\GammaOne$, is in a direct hierarchical relation with a status attribute that is in $\GammaOne$, then that status attribute needs to be a border attribute, to ensure that the hierarchy relation is preserved in the feature.
For instance, stage \textsf{Credit Check} is in $\BCAone$ but not in $\BaseSchema$. Compound stage \textsf{Preliminary Check} is a border attribute, to ensure that hierarchy relation $\textsf{Credit Check} \preceq \textsf{Preliminary Check}$ of $\BCAone$ is included in the extracted feature.
%The constructed feature is B. Then A and C are border attributes.

We now define how a feature is extracted. % from a template and a variant that refines the template.

\begin{deff}[Feature extraction]
	Let $\GammaOne, \GammaTwo$ be two GSM schemas such that $\GammaOne \subseteq \GammaTwo$ (e.g., $\GammaOne$ is the template, $\GammaTwo$ the variant).
	Then $\GammaTwo \setminus \GammaOne$ is the GSM schema $\Gamma=(\Att=\Data \cup \Stg \cup \Mst,\Ev=\EvE \cup \EvC,\preceq,\rules=\rules_+ \cup \rules_-)$ where 
	\begin{itemize}
		%		\item $touch()=\{ a \in \Stg^1 \cup \Mst^1 |  atts(\rules(a)) \cap (\Att^1 \setminus \Att^{1}) \neq \emptyset  \}$
		\item $\Data=\Data^{2} \setminus \Data^{1}$;
		%		\item 
		%		$\Task =(\Task ^{2} \setminus \Task ^{1}) \cup \{ t\in \Task ^{1} | t \in dom(\preceq) \}$;
		\item 
		$\Stg=(\Stg^{2} \setminus \Stg^{1}) \cup (\Stg^{1} \cap borderAtts(\GammaOne,\GammaTwo)) $;
		%		\item 
		%		$\Stg=\Stg^{2} \setminus \Stg^{1}$;
		\item 
		$\Mst=(\Mst^{2} \setminus \Mst^{1}) \cup  (\Mst^{1} \cap borderAtts(\GammaOne,\GammaTwo))$;
		\item $\EvE=\EvE^{2} \setminus \EvE^{1}$;
%		\item 
%		$\Task=\Task^{2} \setminus \Task^{1} $;
		\item
		$\EvC=\EvC^{2} \setminus \EvC^{1}$;
		\item $\preceq = \preceq^{2} \setminus \preceq^{1}$;
		%\item $\rulesPlus=\rulesPlus^{2} \setminus \rulesPlus^{1}$;
%		\item $\tau=\tau^2 \setminus \tau^1$;
		\item for each $a \in \Att$, 
		
		$\rulesPlus(a)=\left\{
		\begin{array}{ll}
		%\parbox{0.6\textwidth}{$\{\varphi[\orig/\psi] ~|~\varphi \in \rulesPlus^{1}(a),\psi \in \rulesPlus^{2}(a)\}  \setminus \{\varphi ~|~\quot{\varphi} \in \rulesPlus^{1}(a)\}$} & \textnormal{, if } a \in \Att^{2} \cap \Att^{1}\\
		\{\varphi[\psi/\orig] ~|~\varphi \in \rulesPlus^{2}(a),\psi \in \rulesPlus^{1}(a)\} & \textnormal{, if } a \in  borderAtts(\GammaOne,\GammaTwo) \\
		%		\{\varphi[\psi/\orig] ~|~\varphi \in \rulesPlus^{2}(a),\psi \in \rulesPlus^{1}(a)\} & \textnormal{, if } a \in \Att \cap \Att^{1}\\
		% & \textnormal{, if } a \in \Att^{2} \cap \Att^{1}\\
		\rulesPlus^{2}(a) & \textnormal{, otherwise}\\
		%	\rulesPlus^{1}(a) & \textnormal{, if } a \in \Att^{1} \setminus \Att^{2}\\
		\end{array}\right.$ \\[1ex]
		%\item for each $a \in \Att$, 
		
		$\rulesMinus(a)=\left\{
		\begin{array}{ll}
		\{\varphi[\psi/\orig] ~|~\varphi \in \rulesMinus^{2}(a),\psi \in \rulesMinus^{1}(a)
		%, \varphi=\psi or \varphi extends \psi 
		\} & \textnormal{, if } a \in  borderAtts(\GammaOne,\GammaTwo) \\
		\rulesMinus^{2}(a) & \textnormal{, otherwise}\\
		\end{array}\right.$
	\end{itemize} 
\end{deff}
Most lines of the definition are straightforward. Above we already explained why border attributes need to be included. For the definition of rules, note that  for a status attribute $a$ in $\GammaTwo$ but not in $\GammaOne$, the rules of $a$ in  $\GammaTwo$ are incorporated in the feature. However, for a status attribute $a$ that is a border attribute, both $\GammaOne$ and $\GammaTwo$ have defined rules. 
In that case, for pairs of rules that are similar, i.e., the rules are equal or the rule in $\GammaTwo$ extends the rule in $\GammaOne$, the feature should contain the  rule of $\GammaTwo$ but with keyword $\orig$ replacing the rule of $\GammaOne$. 
For instance, in $\BaseSchema$ milestone \textsf{PCS} has sentry \textsf{BPECS} while in $\BCAone$ the sentry for \textsf{PCS} is $\textsf{BPECS} \land \textsf{CCS}$. In $\BCAone \setminus \BaseSchema$, the sentry for \textsf{PCS} is $\orig \land \textsf{CCS}$. 
%GammaOne: a<10; on E if x
%GammaTwo: a<10 and y ; on E if x and y
% feature: orig and y; a<10; on E if x; 

%GammaOne: on E if x
%GammaTwo: on E if x ; on E if x and y
% feature: orig and y; orig; 

%GammaOne: on E if x; a<10
%GammaTwo: on E if x ; on E if x and y; a<10 and y
% feature: orig and y; orig;  NO FEATURE possible
%\paragraph{Correctness.}

We state the correctness of the approach with a few lemmas. The first lemma states that feature extraction results in a GSM schema fragment.
\begin{lemma}
	Let $\GammaOne, \GammaTwo$ be two GSM schemas such that $\GammaOne \subseteq \GammaTwo$.
	Then $\Gamma = \GammaTwo \setminus \GammaOne$ is a GSM schema. % if .....
\end{lemma}
%\begin{proof}
%	It is straightforward to check, using Definition \ref{def.border}, that if $a_1 \prec a_2$, both $a_1$ and $a_2$ are in $\Att$ and that each sentry  only references attributes in $\Att$.
%	We need to ensure that for each $(a,b) \in \preceq$, $a, b \in Att$.  $ \subseteq ...$
%\end{proof}
The next lemma ensures that applying a feature that was generated for a template and a variant to the template yields the variant. 
\begin{lemma}
	Let $\GammaOne, \GammaTwo$ be two GSM schemas such that  $\GammaOne \subseteq \GammaTwo$.
	Then $\GammaTwo = (\GammaTwo \setminus \GammaOne) \after \GammaOne$.
\end{lemma}
The next lemma follows easily from the method defined in Fig.~\ref{fig.method}.
\begin{lemma}
	Let $\GammaBase$ be a base GSM schema and $\{\GammaOne,  \upto ,\GammaN\}$ be the set of variant GSM schemas such for each $\Gamma^i$, $\GammaBase \subseteq \Gamma^i$. 
	The set of features generated by the method in Figure \ref{fig.method} is minimal and complete.
	\end{lemma}

%\paragraph{Step 4.} Is it possible that the same feature is constructed multiple times. Yes, consider variants A[T1], B; A[T1], B, C; A[T1], B, C, D. Each time task T1 is added to stage A. The change from A B to A B C to A B C D is orthogonal.
%Therefore, duplicate features are to be removed
%
%No, feature are stored in a set and a set cannot contain duplicates.

% Alternative: input $n$ variant GSM schemas; First step: construct GSM core. 

% Pitfalls: GSM core does not exist, either since the GSM schemas have no common elements, 

% or since the GSM schemas have common elements that are disconnected

% or since the GSM schemas have incompatible hierarchies: A->B->C vs C->B->A
% ---> Impossible because of assumptions that there is a common template from which the variants are derived.

% Pitfall: GSM feature does not exist

%\section{Example}
%We illustrate the method with the running example.
%
%\paragraph{Step 1.}
 
 \section{Evaluation}
 
 To evaluate the feasibility of the approach, we applied the method to a real-world process of an international high tech company with offices in different regions of the worlds. In the process the expired due diligence qualification of a business partner of the company is renewed. The company has defined a standard due diligence process, but offices in certain regions can use their own process variant. 
 
 The standard process and three variants had been modeled before in separate GSM schemas \cite{Yi}. The method could not be applied directly to these GSM schemas, since in one variant a fragment of the standard process was replaced with another fragment.  Therefore, the standard process could not act as base process. We therefore  manually created a base process, specified as GSM schema,  such that both the standard process and the variant refine the  base process. Thus, the standard process becomes another (fourth) variant.  The GSM schemas for all these processes are available in the  appendix.

 %Based on the existing GSM schemas for this process, we defined a base schema $\DDPBaseSchema$ for the standard process and four features that refine the base schema: $\DDPFoneaSchema$, $\DDPFonebSchema$, $\DDPFtwoSchema$, and $\DDPFthreeSchema$; all are available in an online appendix~\cite{Eshuis18}. The first two features are alternatives. Similar to $\FoneSchema$ (Table~\ref{tab.fr1}) and $\FtwoSchema$ (Table~\ref{tab.fr2}), each fragment schema of a feature uses $\orig$ as sentry or as conjunct of a sentry to specify the connection between the base schema and the fragment schema. 
 
 Applying the feature extraction method to the  base process and the four  variants gave the following results.  
In the  variant graph created in step 1, each variant refines the base schema, but not any other variant.  Consequently, in step 2 no transitive edges were removed. 
 In step 3,  four features were created, one for each variant. The GSM schema fragments of the four extracted features are available too in the  appendix.  
  
  \begin{table}[t]
  	\caption{Descriptive statistics of extracted features for Due Diligence Process and its variants\label{tab.case}}
  	\centerline{\small\begin{tabular}{lccccc}\toprule
  			\bf & \bf Base schema & \bf Feature 1 & \bf Feature 2 &\bf Feature 3 &\bf  Feature 4 \\ \midrule
  			\# Non-border stages & 9 & 1 & 2& 1 &1 \\
  			\# Non-border milestones & 15 &1 &2&1&1\\
  			\# Non-border sentries & 60 &5&10&4&5 \\
  			\# Border stages &  & 1 & 1&1&1 \\
  			\# Border milestones & &1 & 1& 0 &0 \\
  			\# Border sentries & & 5 &5&3 &3 \\
  	\end{tabular}}
  \end{table}
  
 Table \ref{tab.case} gives descriptive statistics of the base schema and the four features.
 All extracted features use border attributes to link properly to the base schema. For each border attribute there is at least one sentry that uses the $\orig$ construct. 
 Composing each feature with the base schema gives the original variant. However, since three from the four features are complementary, additional variants can be derived \cite{Eshuis18b}.
 
 This preliminary evaluation shows that the method can be used to extract features from a base schema and a set of GSM schema variants. However, it also shows that some preprocessing can be needed to ensure that all variants refine the base schema. In future work, we plan to extend the method to variants that do not refine the base schema but are overlapping.

%\begin{table}[tb]
%	\caption{Descriptive statistics of extracted features for due Diligence Process and its variants\label{tab.case}}
%	\centerline{\small
%		\begin{tabular}{lC{0.05\textwidth}C{0.05\textwidth}C{0.05\textwidth}C{0.05\textwidth}C{0.05\textwidth}C{0.05\textwidth}C{0.05\textwidth}} %\toprule
%			\bf GSM schema    &
%			\begin{turn}{30}\bf Non-border stages\end{turn}& 
%			\begin{turn}{30}{\bf Non-border milestones}\end{turn}&  
%			\begin{turn}{30}{\bf Non-border sentries}\end{turn}&
%			%\begin{rotate}{90}\# Stages overlap with $\DDPBaseSchema$ \end{rotate}\rule{.5cm}{0pt}&
%			\begin{turn}{30}\bf Border stages\end{turn}&
%			%\begin{rotate}{45}\# Milestones overlap with $\DDPBaseSchema$ \end{rotate}\rule{.5cm}{0pt}&
%			\begin{turn}{30}\bf Border milestones\qquad\end{turn}&
%			\begin{turn}{30}\bf Border sentries\end{turn}
%			\\ \midrule
%			Base schema & \hspace{1ex}9 & 15 &  %30+30=
%			60 \\
%			Feature 1 & 1 & 1 & 5 & 1 & 1 & 5 \\ 
%			Feature 2 &  2 & 2&  10 & 1 & 1 &  5 \\
%			Feature 3 &  1& 1 & 4 & 1 & 0 & 3\\ 
%			Feature 4 & 1& 1&  5 & 1 & 0 & 3 \\ \bottomrule
%	\end{tabular}}
%\end{table}

\section{Related Work}\label{sec.rel}

For artifact-centric process models, there is no directly related work on extraction of  model fragments. The general problem of designing artifact-centric process models, either by defining a methodology for specifying  business artifacts~\cite{Bhattacharya09adata-centric} or by defining an automated synthesis of  artifact-centric process models~\cite{EshuisG16,Fritz09,Lohmann13,PopovaFD15} has been addressed, but without considering fragments that are composed.
%The feature-oriented composition approach facilitates reuse of  model fragments and the generation of different but related  variants, rather than designing a single artifact-centric process model.% from scratch.
%

Alternatives to artifact-centric process models are object-aware~\cite{KunzleR11} and object-centric \cite{ReddingDHI10}  process models and case management models~\cite{AalstWG05,MHPW14,SlaatsMHM13} (though artifact-centric process models can be used for  case management too~\cite{EshuisHY19,Marin}). 
A few of these alternatives support management of process variants \cite{AndrewsSR18} and, related, the use of model fragments \cite{MHPW14,MukkamalaHS13}; we next discuss these in more detail. 

Andrews et al.\ \cite{AndrewsSR18} present concepts for managing variants in object-aware processes. Each object-aware process model is defined by a logged sequence of modeling actions.
A process variant is derived from another process variant by copying the log of modeling actions of that other process variant into a new log and then adding new modeling actions to the log. %updates to a variant are propagated to all derived.
%Updates to a process variant, i.e., new modeling actions added to the log, are propagated to all derived process variants by inserting the new modeling actions in the logs of the derived process variants.
The focus of that paper is on efficiently managing updates for related variants, while this paper focuses on extracting composable fragments from variants.
% such that these variants but also other ones can be composed.
%Variants have hierarchy.
%While this approach supports both fine-grained and coarse-grained composition of process variants, each variant has by definition a unique chain of precursor variants. Whereas feature-oriented composition allows that the same process variant is derived from different, permutable composition chains. 
%Updates to a process variant are propagated to all derived process variants. A feature can be viewed as a sequence of modeling operations that can be applied to multiple variants.

Meyer et al.\ \cite{MHPW14} define an approach for production case management in which procedural, activity-centric process fragments are composed at run-time by linking them, i.e, a case is executed in a distributed fashion by executing linked process fragments. The fragments are linked via shared data objects. 
Mukkamala et al.\ \cite{MukkamalaHS13} define a commutative composition operator on instances of DCR graphs, a declarative, activity-centric process modeling notation. Each DCR graph instance can be viewed as a process fragment being executed.
Both approaches focus on composition of existing process fragments, whereas the approach in this paper focuses on extracting  fragments from variants such that the fragments can be composed.
For activity-centric process models, approaches exist to extract shared fragments in process model repositories \cite{EkanayakeRHF11,WeberRMR11}, to discover configurable process fragments from activity-centric process models \cite{AssyCG15}, or to discover variants from events log  \cite{Jansen-VullersAR06,LiRW}. All these approaches consider graph-like process models, which differ considerably from  declarative, rule-based process models like GSM schemas.
%Other works consider mining of declarative activity-centric process models, but do not consider discovery of process fragments \cite{ChesaniLMMRS09,MaggiMA11}.

In software engineering, feature extraction has been studied for software artifacts, e.g.\ \cite{LinsbauerLE17,MartinezZBKT15}. However, those features need to be manually identified, whereas in our method features are derived automatically. Studying how domain knowledge from experts can improve the quality of the generated features is an interesting direction for further work.
%To the best of our knowledge, there is no related work that discovers  fragments from declarative  process models.

In sum, the main contribution of this paper is an approach to extract  composable fragments from declarative, artifact-centric process variants.

\section{Conclusion}\label{sec.conc}

This paper has defined a novel approach to extract model fragments, viewed as features, from declarative, artifact-centric process model variants.  Using feature composition \cite{Eshuis18b} the declarative fragments can be composed in a declarative way into the original variants, but also other variants can be composed. 
%A key challenge has been to reconcile features, which are additive, with modifications of variants, which are non-additive, since they may be the result of deletions~\cite{EshuisHY19}.
The approach can be used to decompose variants of case management templates into reusable fragments, that encode well-known modifications. This way, complex case management variants  can be  efficiently composed in a declarative way.  

There are several  directions for future work. An open challenge is to reconcile features, which are additive, with modifications of variants, which are non-additive, since they may be the result of deletions~\cite{EshuisHY19}.
%Feature composition is additive, each feature adding but not deleting stages, milestones or sentries. Therefore, GSM schema variants that were derived by deleting stages, milestones and sentries must be handled in a different way. We plan to tackle this in future work.
Next, we plan to realize a tool implementation of the approach geared towards  CMMN \cite{OMG:CMMN:Beta1}. In addition, we plan to apply this tool to several case study examples to further evaluate the approach.
 
\bibliographystyle{splncs03}
%\bibliography{t}

\newpage
\section*{Appendix}

\def\DDPBaseSchema{{\textsf{\small DDP}}_{base}}
\def \ModSchema {{\textsf{\small BCA}}^{alt\_check}}
\def \CCSchema {{\textsf{\small BCA}}^{cred\_check}}
\def \AMCSchema {{\textsf{\small BCA}}^{amc\_check}}
\def\DDPFoneaSchema{{\textsf{\small F}}^{\sf DDP}_{1}}
\def\DDPFonebSchema{{\textsf{\small F}}^{\sf DDP}_{2}}
\def\DDPFtwoSchema{{\textsf{\small F}}^{\sf DDP}_{3}}
\def\DDPFthreeSchema{{\textsf{\small F}}^{\sf DDP}_{4}}

\def\DDPoneaSchema{{\textsf{\small DDP}}_{1}}
\def\DDPonebSchema{{\textsf{\small DDP}_{2}}}
\def\DDPtwoSchema{{\textsf{\small DDP}_{3}}}
\def\DDPthreeSchema{{\textsf{\small DDP}_{4}}}

This appendix describes features extracted from the variant GSM schemas presented in chapter 5 of Yi~\cite{Yi}.
The GSM schemas of Yi have been converted into the notation used in this paper. Each GSM schema models three interacting artifacts. To suit the single artifact framework used in the main text, we have converted the multi-artifact schemas of Yi into single-artifact schemas.
Another change is that Yi models a base process that contains a GSM fragment that in one variant is replaced with another GSM fragment.  To be able apply the feature extraction method, we have modeled a new base process that contains neither of these fragments. The new base process is refined by the old base process and all the variants. % both fragments as separate, alternative features. The base process of Yi corresponds to the base schema and feature 1a, both depicted below.

\section*{Base schema}

The  (new) base schema concerns the process of due diligence qualification of business partners for an international hightech company that has international offices in different countries in Europe, Asia and America. The process starts when a due diligence qualification of a business partner has expired. To renew the qualification, first information needs to be collected about the partner such that the company has a sufficient level of information (milestone \textsf{\small Confirmed}). Next, based on the collected information, the due diligence qualification is checked. In some cases, mitigation actions need to be taken for the RequestForQuotes that the business partner can receive from the company.  Finally, the due diligence qualification is signed off, resulting in either an approved or rejected status. An approved due diligence leads to an activated partner status, a rejected due diligence leads to an inactivated partner status.

\begin{figure}[ptb]
	\centerline{
		\includegraphics[scale=0.45]{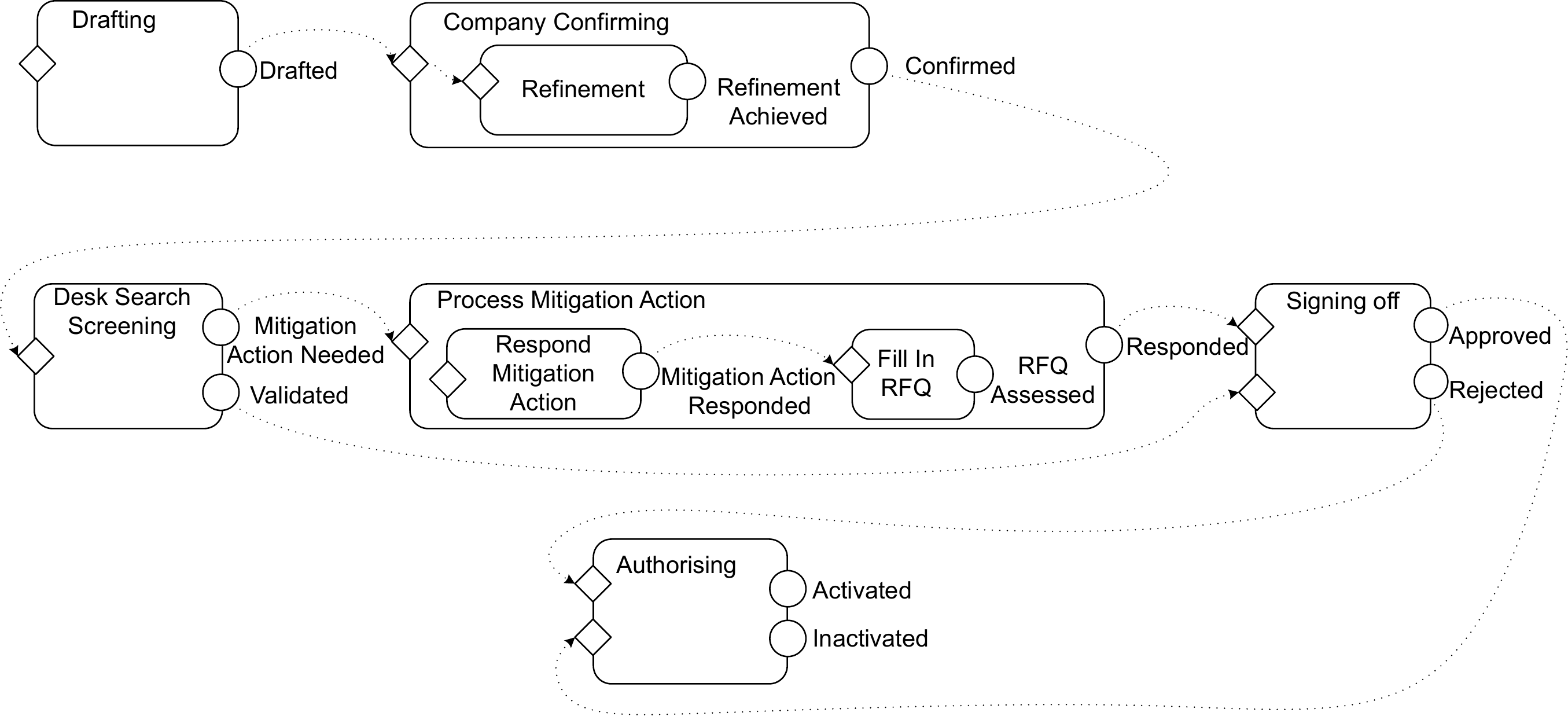}}
	\caption{\label{fig:ddp} Base GSM schema Due Diligence Process ($\DDPBaseSchema$)}
\end{figure}

Figure~\ref{fig:ddp} shows the GSM schema $\DDPBaseSchema$ and Table~\ref{tab.ddp.stg} and~\ref{tab.ddp.mst} the sentries of the stages and milestones, respectively.

\begin{table}[tbh]
	\caption{Stages and sentries  for $\DDPBaseSchema$\label{tab.ddp.stg}. ``;" separates different sentries}
	\centerline{\footnotesize\sffamily
		\begin{tabular}{p{0.22\linewidth}p{0.35\linewidth}p{0.4\linewidth}}
			\toprule
			{\bf Stage} & \bf Plus sentries (guards) & \bf Minus sentries (closing) \\ \midrule
			{\sf Drafting} & \sf \evt{RegularDDQRenewal} &{\sf +Drafted}  \\
			{\sf CompanyConfirming} & {+Drafted} & +Confirmed; +Drafting\\
			%			{\sf FirstVerification}  & +CompanyConfirming & +Verified; -CompanyConfirming\\		
			Refinement & CompanyConfirming  & +RefinementVersionAchieved ; -CompanyConfirming\\
			DeskSearchScreening & +Confirmed& +MitigationActionNeeded;+Validated\\
			ProcessMitigatio\-Action & +MitigationActionNeeded & +Responded;-DeskSearchScreening\\
			RespondMitigation\-Action & +ProcessMitigationAction & +MitigationActionResponsed; -ProcessMitigationAction\\
			FillinRFQ & MitigationActionNeeded $\land$ ProcessMitigationAction & +RFQassessed; +Respond MitigationAction; -ProcessMitigationAction\\
			SigningOff  & Responded; +Validated & +Approved; +Rejected; +ProcessMitigation\\
			Authorizing & +Approved;+Rejected &+Activated; -Inactivated\\
		\end{tabular}		
	}		
\end{table}

\begin{table}[tb]
	\caption{Milestones and sentries for $\DDPBaseSchema$\label{tab.ddp.mst}.``;" separates different sentries}
	\centerline{\footnotesize \sffamily
		\begin{tabular}{p{0.25\linewidth}p{0.35\linewidth}p{0.3\linewidth}}
			\toprule
			\bf Milestone  & \bf Plus sentries (achieving) & \bf Minus sentries (invalidating)  \\ \midrule	
			Drafted & \cmpltn{Drafting} & +Drafting\\		
			%			Verified & \cmpltn{FirstVerification} $\land$ CompanyConfirming & +RespondMitigationAction\\
			RefinementVersion\-Achieved & \cmpltn{PhilipsRefinement}  $\land$ CompanyConfirming & +Refinement\\
			Confirmed & +RefinementVersionAchieved & +CompanyConfirming\\
			MitigationAction\-Needed & mitigation\_needed & +Validated; +DeskSearchScreening\\
			Validated & $\lnot$ mitigation\_needed & +MitigationActionNeeded; +DeskSearchScreening \\
			MitigationAction\-\textsl{}Responded & \cmpltn{RespondMitigationAction} $\land$ ProcessMitigationAction & +FillinRFQ\\
			RFQassessed & \cmpltn{FillinRFQ} $\land$ ProcessMitigationAction & +FillinRFQ\\
			Responded & +RFQassessed  & +ProcessMitigationAction\\
			Approved & signed\_off & +Rejected; +Signing off \\
			Rejected & $\lnot$ signed\_off & +Approved; +Signing off \\
			Activated & partner\_status\_activated & +Inactivated; +Authorizing\\
			Inactivated & $\lnot$ partner\_status\_activated & +Activated; +Authorizing\\
		\end{tabular}
	}
\end{table}

\clearpage
%The next two variants are alternatives.

\section*{Variant 1} Variant $\DDPoneaSchema$
specifies that a first verification check is performed before stage \textsf{\small Refinement}; see Table~\ref{tab.ddp.1a} for the new stages and milestones, plus the stages and milestones from the base schema whose sentries have been modified in this variant.
\begin{table}[h]
	\caption{New and modified sentries  for variant  $\DDPoneaSchema$. S=Stage; M=Milestone }\label{tab.ddp.1a}
	\centerline{\footnotesize\sffamily
		\begin{tabular}{p{0.07\linewidth}p{0.225\linewidth}p{0.3\linewidth}p{0.4\linewidth}}\small
			\bf Type & \bf Name & \bf Plus sentries & \bf Minus sentries \\ \hline
			% 	\textnormal{S}&		{\sf CompanyConfirming} &  {+Drafted} & +Confirmed; +Drafting \\
			\textnormal{S}&		{\sf FirstVerification}  & +CompanyConfirming & +Verified;
			-CompanyConfirming\\	
			\textnormal{M}&	Verified & \cmpltn{FirstVerification} $\land$ CompanyConfirming & +FirstVerification\\
			\textnormal{S}&	 Refinement & CompanyConfirming $\land$ Verified   & +RefinementVersionAchieved ; -CompanyConfirming ; +FirstVerification\\
	\end{tabular}}
\end{table}

%\paragraph{Feature 1a.}

Feature $\DDPFoneaSchema=\DDPoneaSchema \setminus \DDPBaseSchema$ ; see Table~\ref{tab.ddp.fr1a}.

\begin{table}[h]
	\caption{Sentries for  feature  $\DDPFoneaSchema$. S=Stage; M=Milestone }\label{tab.ddp.fr1a}
	\centerline{\footnotesize\sffamily
		\begin{tabular}{p{0.07\linewidth}p{0.225\linewidth}p{0.3\linewidth}p{0.4\linewidth}}\small
			\bf Type & \bf Name & \bf Plus sentries & \bf Minus sentries \\ \hline
			\textnormal{S}&		{\sf CompanyConfirming} & \orig & \orig \\
			\textnormal{S}&		{\sf FirstVerification}  & +CompanyConfirming & +Verified;
			-CompanyConfirming\\	
			\textnormal{M}&	Verified & \cmpltn{FirstVerification} $\land$ CompanyConfirming & +FirstVerification\\
			\textnormal{S}&	 Refinement & Verified $\land$ \orig  & \orig ; +FirstVerification\\
	\end{tabular}}
\end{table}

\clearpage
\section*{Variant 2} Variant $\DDPonebSchema$
specifies that a PreCheck and ExpertReview are performed before stage \textsf{\small Refinement}; see Table~\ref{tab.ddp.1b} for the new stages and milestones, plus the stages and milestones from the base schema whose sentries have been modified in this variant.

\begin{table}[h]
	\caption{New and modified sentries  for variant  $\DDPonebSchema$. S=Stage; M=Milestone }\label{tab.ddp.1b}
	\centerline{\footnotesize\sffamily
		\begin{tabular}{p{0.07\linewidth}p{0.225\linewidth}p{0.3\linewidth}p{0.4\linewidth}}\small
			\bf Type & \bf Name & \bf Plus sentries & \bf Minus sentries \\ \hline
			%		\textnormal{S}&		{\sf CompanyConfirming} & \orig & \orig \\
			\textnormal{S}&		{\sf PreCheck}  & +CompanyConfirming & +Checked; -CompanyConfirming\\	
			\textnormal{S}&		{\sf ExpertReview}  & Checked $\land$ CompanyConfirming & +Reviewed; -CompanyConfirming\\	
			\textnormal{M}&	Checked & \cmpltn{PreCheck} $\land$ CompanyConfirming & +PreCheck\\
			\textnormal{M}&	Reviewed & \cmpltn{ExpertReview} $\land$ CompanyConfirming & +ExpertReview\\
			\textnormal{S}&	 Refinement & CompanyConfirming $\land$ Reviewed   & +RefinementVersionAchieved ; -CompanyConfirming ; +PreCheck\\
	\end{tabular}}
\end{table}

%\section*{Feature 1b.}

Feature $\DDPFonebSchema=\DDPonebSchema \setminus \DDPBaseSchema$ ; see Table~\ref{tab.ddp.fr1b}.

\begin{table}[h]
	\caption{Sentries for partial feature  $\DDPFonebSchema$. S=Stage; M=Milestone }\label{tab.ddp.fr1b}
	\centerline{\footnotesize\sffamily
		\begin{tabular}{p{0.07\linewidth}p{0.225\linewidth}p{0.3\linewidth}p{0.4\linewidth}}\small
			\bf Type & \bf Name & \bf Plus sentries & \bf Minus sentries \\ \hline
			\textnormal{S}&		{\sf CompanyConfirming} & \orig & \orig \\
			\textnormal{S}&		{\sf PreCheck}  & +CompanyConfirming & +Checked; -CompanyConfirming\\	
			\textnormal{S}&		{\sf ExpertReview}  & Checked $\land$ CompanyConfirming & +Reviewed; -CompanyConfirming\\	
			\textnormal{M}&	Checked & \cmpltn{PreCheck} $\land$ CompanyConfirming & +PreCheck\\
			\textnormal{M}&	Reviewed & \cmpltn{ExpertReview} $\land$ CompanyConfirming & +ExpertReview\\
			\textnormal{S}&	 Refinement & Reviewed $\land$ \orig  & \orig ; +PreCheck\\
	\end{tabular}}
\end{table}

\clearpage
\section*{Variant 3}

Variant $\DDPtwoSchema$ specifies that for a new partner the completion of the profile triggers the due diligence qualification process; see Table~\ref{tab.ddp.2}  for the new stages and milestones, plus the stages and milestones from the base schema whose sentries have been modified in this variant.

\begin{table}[h]
	\caption{New and modified sentries  for variant  $\DDPtwoSchema$. S=Stage; M=Milestone }\label{tab.ddp.2}
	\centerline{\footnotesize\sffamily
		\begin{tabular}{p{0.07\linewidth}p{0.225\linewidth}p{0.45\linewidth}p{0.2\linewidth}}\small
			\bf Type & \bf Name & \bf Plus sentries & \bf Minus sentries \\ \hline
			\textnormal{S}&		{\sf InitiatePartner} & \evt{NewPartnerRequest} & +ProfileCreated \\
			\textnormal{M}&	ProfileCreated & \cmpltn{InitiatePartner} & +InitiatePartner\\
			\textnormal{S}&	 Drafting & \sf \evt{RegularDDQRenewal}; ProfileCreated &{\sf +Drafted} \\
	\end{tabular}}
\end{table}

%\paragraph{Feature 2.}

Feature $\DDPFtwoSchema=\DDPtwoSchema\setminus \DDPBaseSchema$ ; see Table~\ref{tab.ddp.fr2}.

\begin{table}[h]
	\caption{Sentries for partial feature  $\DDPFtwoSchema$. S=Stage; M=Milestone }\label{tab.ddp.fr2}
	\centerline{\footnotesize\sffamily
		\begin{tabular}{p{0.07\linewidth}p{0.225\linewidth}p{0.3\linewidth}p{0.3\linewidth}}\small
			\bf Type & \bf Name & \bf Plus sentries & \bf Minus sentries \\ \hline
			\textnormal{S}&		{\sf InitiatePartner} & \evt{NewPartnerRequest} & +ProfileCreated \\
			\textnormal{M}&	ProfileCreated & \cmpltn{InitiatePartner} & +InitiatePartner\\
			\textnormal{S}&	 Drafting & \orig; ProfileCreated  & \orig\\
	\end{tabular}}
\end{table}

\clearpage
\section*{Variant 4}

Variant $\DDPthreeSchema$ specifies that the due diligence process can be completed in a fast way; see Table~\ref{tab.ddp.3}  for the new stages and milestones, plus the stages and milestones from the base schema whose sentries have been modified in this variant.

\begin{table}[h]
	\caption{New and modified sentries for variant  $\DDPthreeSchema$. S=Stage; M=Milestone }\label{tab.ddp.3}
	\centerline{\footnotesize\sffamily
		\begin{tabular}{p{0.07\linewidth}p{0.25\linewidth}p{0.35\linewidth}p{0.3\linewidth}}\small
			\bf Type & \bf Name & \bf Plus sentries & \bf Minus sentries \\ \hline
			\textnormal{S}&		{\sf ConfirmNoScreening} & \evt{FastDDQRenewal} & +NoScreeningConfirmed \\
			\textnormal{M}&	NoScreeningConfirmed & no\_screen\_authorization & $\lnot$no\_screen\_authorization; +ConfirmNoScreening\\
			\textnormal{S}&	 SigningOff & Responded; +Validated; NoScreeningConfirmed  & +Approved; +Rejected; +ProcessMitigation\\
	\end{tabular}}
\end{table}

%\paragraph{Feature 3.}

Feature $\DDPFthreeSchema$ specifies that the due diligence process can be completed in a fast way; see Table~\ref{tab.ddp.fr3}.

\begin{table}[h]
	\caption{Sentries for partial feature  $\DDPFthreeSchema$. S=Stage; M=Milestone }\label{tab.ddp.fr3}
	\centerline{\footnotesize\sffamily
		\begin{tabular}{p{0.07\linewidth}p{0.25\linewidth}p{0.35\linewidth}p{0.35\linewidth}}\small
			\bf Type & \bf Name & \bf Plus sentries & \bf Minus sentries \\ \hline
			\textnormal{S}&		{\sf ConfirmNoScreening} & \evt{FastDDQRenewal} & +NoScreeningConfirmed \\
			\textnormal{M}&	NoScreeningConfirmed & no\_screen\_authorization & $\lnot$no\_screen\_authorization; +ConfirmNoScreening\\
			\textnormal{S}&	 SigningOff & \orig; NoScreeningConfirmed  & \orig\\
	\end{tabular}}
\end{table}

%\paragraph{Feature dependencies.}
%
%Features $\DDPFoneaSchema$ and $\DDPFonebSchema$ are alternative to one another. The other features are complementary.
%

\end{document}